\documentclass{gPAA2e}
\usepackage{amssymb}
\usepackage{amsmath}
\usepackage{color}

\begin{document}



\title{The \textit{mise en sc\'{e}ne} of memristive networks:\\ effective memory, dynamics and learning }

\author{Francesco Caravelli$^{\rm a}$$^{\ast}$\thanks{$^\ast$Corresponding author. Email: fc@lims.ac.uk
\vspace{6pt}}\\\vspace{6pt}  $^{\rm a}${\em{Invenia Labs, 27 Parkside Place, Cambridge CB1 1HQ, UK\\London Institute of Mathematical Sciences, 35a South Street, London W1K 2XF, UK}}}

\maketitle

\begin{abstract}
We discuss the properties of the dynamics of purely memristive circuits using a recently derived consistent equation for the internal memory variables of the involved memristors. In particular, we show that the number of independent memory states in a memristive circuit is constrained by the circuit conservation laws, and that the dynamics preserves these symmetries by means of a projection on the physical subspace. Moreover, we discuss other symmetries of the dynamics under various transformations of the internal memory, and study the linearized and strongly non-linear regimes of the dynamics. In the strongly non-linear regime, we derive a conservation law for the internal memory variables. We also provide a condition on the reality of the eigenvalues of Lyapunov matrices describing the linearized dynamics close to a fixed point. We show that the eigenvalues ca be imaginary only for mixtures of passive and active components. Our last result concerns the weak non-linear regime. We show that the internal memory dynamics can be interpreted as a constrained gradient descent, and provide the functional being minimized. This latter result provides another direct connection between memristors and learning. 

\bigskip

\bigskip
\end{abstract}



\maketitle

\section{Introduction}

There has been increasing interest in the properties of networks with memory. In the field of complex networks, memory is emerging as a new direction of study \cite{Caravelli2015,Caravelli2016} in order to understand the properties of dynamical networks.  

Memristors in particular have been attracting a renewed interest as these devices 
 resemble swarms in solving certain optimization problems \cite{Dorigo,traversa14b,Traversa2014,Traversa2015}. Memristors are 2-port devices which behave as resistances that change their values as a function of current or voltage. This type of memory is a common feature in many physical systems \cite{pershin11a} and thus of general interest.
Moreover, memristors have been proposed as building blocks for unconventional (or neuromorphic) computing \cite{indiveri,traversa13a}, given that they are becoming easier to fabricate \cite{Avizienis,Stieg12}, although in specialized laboratories. It is thus interesting to study the behavior of circuits of memristors, which we call memristive networks. These can serve also as simple models for further understanding the collective behavior and learning abilities of many biological systems \cite{pershin09b,diventra13a,pershin11d,festchua,pershin13b}, including the brain \cite{chialvo,pershin10c} and its critical aspects.  The behavior of memristors is in spirit similar also to slime molds \cite{adamatzkym}.

In a recent paper \cite{Caravelli2016rl}, a derivation of a ``inner memory'' differential equation for purely memristive networks was obtained. It has been shown that several phenomena can be derived using this equation, such as a slow relaxation in the DC-controlled case, and an approximate analytical solution in the AC-controlled case. In order to derive such an equation, several graph-theoretic tools were used, which inherently showed the underlying freedom in describing the dynamics of the memory.
In this paper, we further study such an equation and its underlying properties \cite{strukov08, strukov05a, chua76a, diventra09a, stru12, demin15}. As an example, we provide an exact solution for the simple case of series of memristors in the \textit{mean field} approximation, showing that it matches with the solution derived by simple circuit analysis.
We then study the backbones of the dynamics: how the constraints structure typical of linear circuits is inherited by memristive networks dynamics. We describe what these constraints imply for the effective independent memory states. Also, we study the properties of the equation, its symmetries and variable transformations, and the weak and strong non-linear regimes. 

We are moreover able to prove a constraint on the presence of oscillations around fixed points of the dynamics in the case in which there are no mixtures of passive and active elements. To conclude, we show in the limit of weak non-linearity that memristors perform ``learning": we cast the dynamics of the internal memory as constrained gradient descent, and provide the functional being minimized. Conclusions follow.

\section{Memristive circuits}
\subsection{Memristive networks}
\begin{figure}
\centering
\includegraphics[width=0.3\textwidth,bb=0 0 400 400]{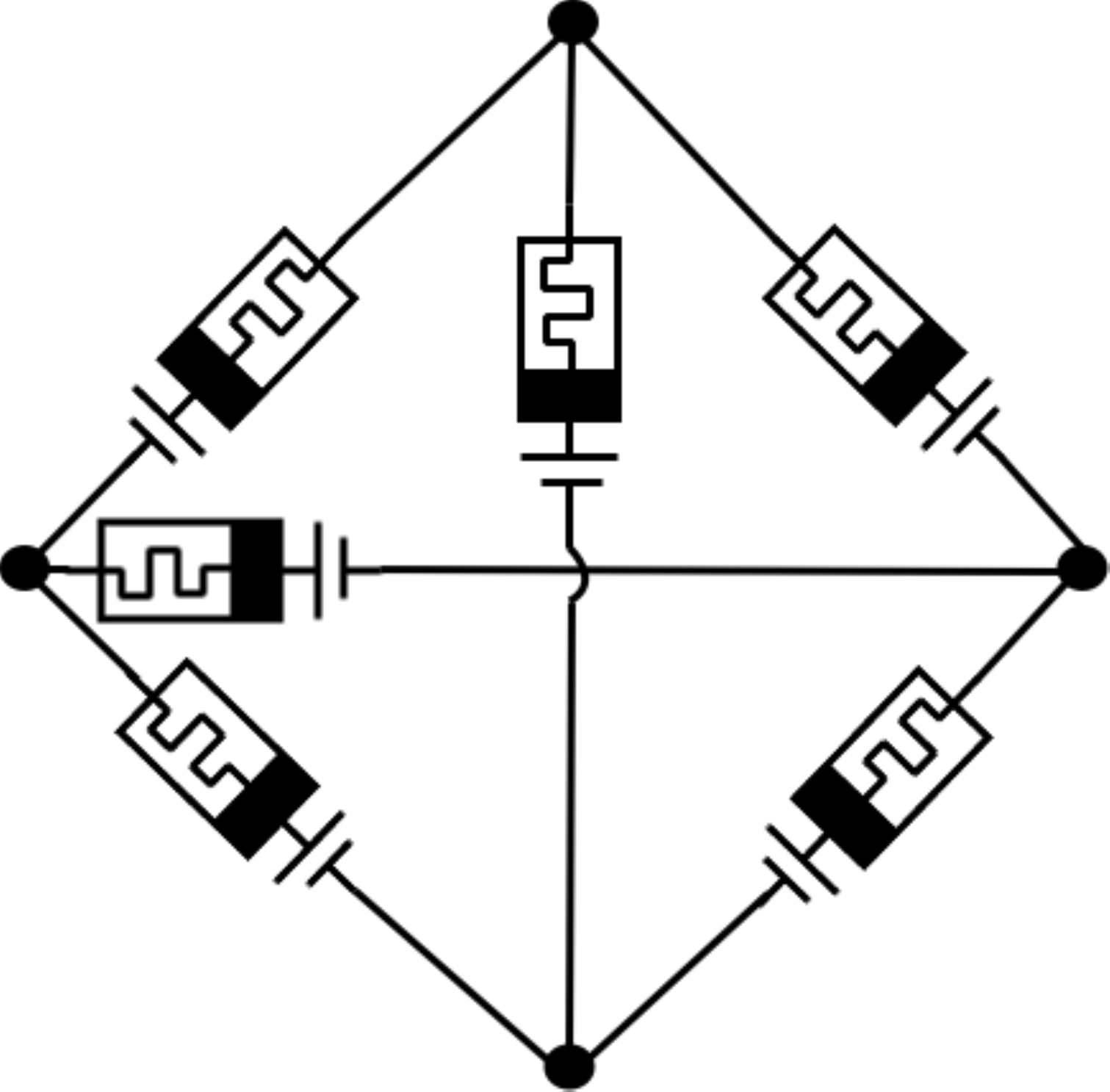}
\caption{An example of memristive network considered in this paper. Given a graph $\mathcal G$, each edge of the graph can be associated with a series of a voltage generator and a memristor.}
\label{fig:circ}
\end{figure}

We begin by briefly introducing the type of circuits we are interested in. First of all, we consider a particular class which, given a graph $\mathcal G$ associated to the topology of the circuit, \textit{each} edge of the graph can be replaced by a series of a voltage generator and a memristor. Thus, we do not consider the case in which inductors, resistors or capacitors are present in the circuit (although nothing obstruct a generalization which includes these components). An example of such circuit is provided in Fig. \ref{fig:circ}, where the graph $\mathcal G$ is a complete graph $K_4$. Also, we consider the case in which each memristor has a resistance which varies linearly as a function of the internal parameter $w$, e.g.
\begin{equation}
R(w)=R_{off} w+(1-w) R_{on},
\label{eq:memristor}
\end{equation}
 where $R_{off}\geq R_{on}$ are the two limiting resistances of the memristor, and $0\leq w \leq 1$.
We also consider a simple dynamics for the internal parameter $w$, which we identify as ``internal memory":
\begin{equation}
\frac{d}{dt} w=\alpha w-j \frac{R_{on}}{\beta} I.
\label{eq:dynsing}
\end{equation}
In eqn. (\ref{eq:dynsing}), $\alpha$ and $\beta$ are the variables which set the timescales of the decay and the reinforcement due to the flow of the current respectively. The constant $j=\pm 1$ is called polarity and is associated to the response (increasing or decreasing resistance) of the memristor to an external potential. Although here we describe only the theoretical properties of the dynamics, this type of memristors can be experimentally realized using atomic switches, which are based on silver ions filaments  \cite{AtomicSwitch1, AtomicSwitch2}. This type of memristor is called \textit{ideal}, and it satisfies the current-voltage relationship typical of a resistor: $V=R(w) I$. Thus, it has the zero-crossing property, i.e. $V=0$ if $I=0$ and viceversa.
Let us make a few comments. We consider the case of a voltage generator in series with the memristor simply because in the parallel case the dynamics of each specific memristor is trivial, and depends only on the voltage applied by the generator added in parallel. Moreover, the notation of eqn. (\ref{eq:memristor}) differs from the one originally introduced in \cite{strukov08} by a transformation $w\rightarrow 1-w^\prime$. However, this change can be reabsorbed in the definition of polarity of each memristor, as $\frac{d}{dt} w=-\frac{d}{dt} w^\prime$ in eqn. (\ref{eq:dynsing}). 
We favor the notation of eqn. (\ref{eq:memristor}) as the limiting internal variable value $w=0$ corresponds to the lower resistive state $R_{on}$. Physical memristors \cite{AtomicSwitch1} satisfy a relaxation into the maximum resistance value $R_{off}$. We thus need to be careful, as in our notation $\alpha>0$ corresponds to a relaxation at zero potential to a $R_{off}$ (insulating) state. Physically, this relaxation is related to an Ostwald ripening phenomenon \cite{AtomicSwitch2}.

Another point that we want here to make, is the distinction between active and passive components in our formalism. In the present paper, a passive component is an element which has the characteristics of a \textit{positive resistance}.
On the other hand, an \textit{active} component is interpreted as a \textit{negative resistance}, i.e. it satisfies $V=-R I$.
\subsection{Graph theory, circuits and memristors}\label{sec:graphs}
In the previous section we have introduced the simplest ideal memristor which will be considered in the bulk of this paper.
In this section we recall the basic notions of graph theory which were used to derive the consistent internal memory differential equation in \cite{Caravelli2016rl,nilsson,bollobas}, which is the starting points for the analysis which follows. First of all, we consider a graph $\mathcal G$ (a circuit) with $N$ nodes and $M$ edges (memristors) which describes the connectivity properties of the circuit. It is standard to start by choosing an orientation $\mathcal O$ for the currents flowing in the circuit but, as we will see later, the analysis is independent from this choice from a physical point of view. 
In order for the graph to represent a circuit, the graph must be connected and the degree of each node $i$ must satisfy $d_i>2$, meaning each node is attached to at least two edges. 

For the sake of clarity, we use latin indices for the edges, and greek indices for the nodes; greek indices with tildes will represent instead cycles on the graph. For instance,  we will introduce a potential vector $p_\alpha$, and for each edge a current $i_k$.  

\begin{figure}
\centering
\includegraphics[width=0.3\textwidth,bb=0 0 400 400]{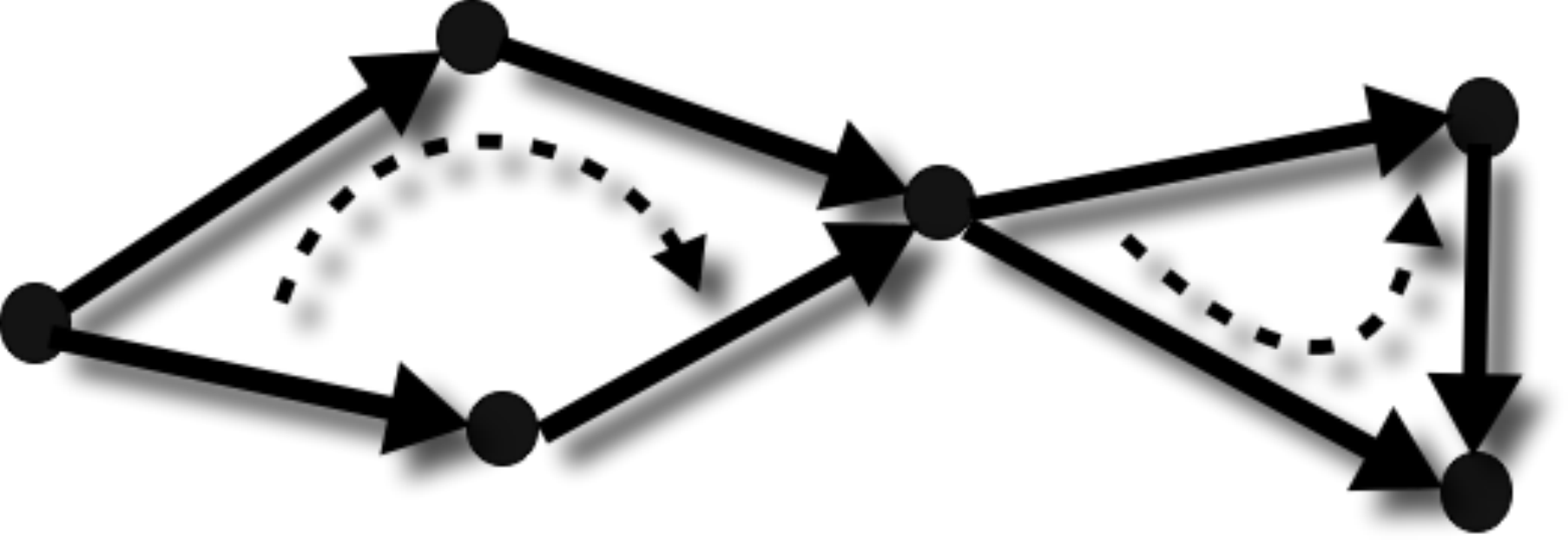}
\caption{Example of fully oriented network: an orientation for each edge and for each cycle was made.}
\label{fig:orchoice}
\end{figure}

We now introduce a few mathematical definitions in order to clarify the discussion. Once an orientation $\mathcal O$ has been assigned, and a set of oriented cycles is obtained, we can introduce two key matrices which will be used in the following: the directed incidence matrix $B^{\mathcal O}_{\alpha k}$, which is a matrix of size $N \times M$, and the cycle matrix $A^{\mathcal O}_{\tilde \xi \beta}$, which is of size $C \times M$, where $C$ is the number of cycles of the graph, $M$ the number of edges and $N$ the number of nodes. The incidence matrix has an entry $-1$ if an (oriented) edge is leaving a node, $+1$ if it is incoming to a node, and $0$ otherwise. The directed incidence matrix $B_{\alpha k}$ labels edges on the rows and nodes on the columns: $B_{\alpha k}$ takes values $+1$ is an edge $\alpha$ is incoming on a node $k$, $-1$ if it is outgoing, and $0$ if the two are not incident. The cycle matrix has loop labels on its columns and edges on the rows:  $A_{\tilde \alpha \beta}$ has entry $-1$ if the directed edge $\beta$ is in the opposite direction of a chosen cycle $\tilde \alpha$, $+1$ if it shares the same orientation, and $0$ if it does not belong to that cycle. 
In what follows, we will assume that an orientation for the cycles and the currents have been chosen, as in Fig. \ref{fig:orchoice}. 

One thing that should be stressed is that  $B B^t$ and $B^t B$ are very different operators (where $\ ^t$ represents the matrix transpose): the former is usually called \textit{laplacian} and is a matrix which acts on the set of \textit{nodes}, meanwhile the latter is usually called  \textit{edge laplacian} \cite{edgelap1,edgelap2} and acts on the set of \textit{edges}. Both operators are however positively defined, as $\vec e \cdot B^t B \vec e=(B\vec e) \cdot (B \vec e)\geq 0$, and in the other case  $\vec n \cdot B B^t \vec n=(B^t\vec n) \cdot (B^t \vec n)\geq 0$.

The conservation of the current at each node, the \textit{first Kirchhoff law} or Current Law (KCL), can be written in terms of the incidence matrix $B$ as $\sum_{j=1}^M B_{\alpha j} i_j=B \vec i=0$. This set of equations contains a superfluous one. Thus, in order for $B$ to have the linear independence of the rows, it is common practice to remove one of the rows and work with the \textit{reduced incidence matrix} $\tilde B$. In the following, we will thus consider only results derived with this matrix rather than the full one and remove the $\tilde {\ }$. \footnote{For the interested reader, we note that in the language of \textit{discrete cohomology} \cite{cohom1}, the incidence matrix represents the \textit{boundary} operator $\partial \cdot$. Such representation exists for any oriented graph. }
The incidence matrix can also be used to calculate the voltage applied to each resistor from the potential at the nodes.
Given a potential vector based on the nodes $\vec p=\{p_\xi\}$, the vector of voltages applied to each resistor can be written as $\{\bar v\}_k=v_k= \sum_{\xi} B_{ \xi k}^t p_\xi$.

 As for the case of the incidence matrix, also in the case of the cycle matrix one has to consider the reduced cycle matrix $\tilde A$ when one row has been removed.
 The \textit{second Kirchhoff law}, or Voltage Law (KVL) can be expressed mathematically  as $\sum_\beta A_{\tilde \xi\beta} v_\beta=0$. This implies that the voltage on each loop (or  \textit{mesh} in circuits) must be zero, which can also be written as 
$\sum_\beta A_{\tilde \xi k} R_\beta i_\beta=0$. It is possible to write this alternatively as
$\sum_\beta A_{\tilde \xi \beta} \sum_{\xi} B_{ \xi \beta}^t p_\xi=0$. Since this is true for arbitrary potential vectors $p_\xi$, this implies that in general one has $B\cdot A^t=A\cdot B^t\equiv 0$. Effectively, this  equation represents the conservation of energy, which in the language of circuits is called \textit{Tellegen's theorem}. 

There are two ways to construct the basis of a circuit: either by using the tree and cycles decomposition, or alternatively by using the chords or the co-chords decomposition \cite{polettini}. Here we consider the standard co-chords decomposition, which is based on \textit{spanning trees}. Let us first introduce a maximal spanning tree $\mathcal T$, whose elements are called \textit{co-chords}, and the set of edges of the graph not included in the tree, \textit{or chords}, are given by $\bar {\mathcal T}$.  If there is more than one tree, each tree has the same cardinality. Each chord element, $\bar {\mathcal T}$, can be assigned to a cycle, called \textit{fundamental loop}. The number of fundamental loops is constrained by the topology of the circuit, and is given by $L=M-N+1$: this is the number of edges minus the cardinality of the tree, $|\mathcal T|=N-1$.  We resort to the appendix of \cite{Caravelli2016rl} for all the details of the derivation of the equation for the internal memory dynamics. The important fact is however that using the Kirchhoff laws, it is possible to derive the following exact solution for the circuit currents, based only on the source vector $\vec S(t)$, the resistance matrix $R$ and the cycle matrix \cite{bollobas}:
\begin{equation}
\vec i=A^t {\vec i}_c=- A^t(A R A^t)^{-1} A \vec S(t). 
\label{eq:init0}
\end{equation}
For the case of linear memristors, i.e. $R(w)=R_{on} (1-w)+ w R_{off}$,  we have shown in \cite{Caravelli2016} that it is possible to rearrange the right hand side in terms of only a projector operator $\Omega=A^t(A A^t)^{-1} A$ on the space of cycles. This is done by carefully inverting only matrices which are invertible. The vector $\vec S$ represents the voltage source in series to the memristances and is a vector with a number of entries equal to the number of memristors. This is due to the fact that in our definition each memristor is in series with a voltage generator.
 Since $A$ is a reduced incidence matrix, then $AR A^t$ (which is usually called \textit{augmented cyclic matrix}) is always invertible for non-zero (thus also negative) resistances \cite{riaza}. 

 Specifically, in \cite{Caravelli2016rl} the following exact equation for the internal memory of an ideal memristive  circuit was derived:
\begin{eqnarray}
\label{eq:exacteqc}
\frac{d \vec W}{dt}&=&\alpha \vec W-\frac{\mathcal J}{\beta}\left(I+\xi \bar \Omega W\right)^{-1} \bar \Omega \bar S,
\label{eq:exacteq} 
\end{eqnarray}
where $\bar S=P \vec S$ and  $\bar \Omega=A^t(  AP  A^t)^{-1} AP=A^t(  \bar A A^t)^{-1} \bar A$.\footnote{
Such equation can be generalized to the case in which the internal memory of each memristor changes linearly in the voltage, rather than the current. The result was obtained in \cite{Caravelli2016rl} and reads:
\begin{eqnarray}
\frac{d \vec W}{dt}&=&\alpha \vec W-\frac{1}{\beta}\left(I+\xi W\right)\left(I+\xi \bar \Omega W\right)^{-1} \bar \Omega \bar S. 
\end{eqnarray}
} $P$ and $J$ are diagonal matrices made of only $\pm1$. In the case of the matrix $P$, elements associated with $-1$ are active elements (negative resistance), meanwhile elements with $+1$ are passive (positive resistance). The matrix $\mathcal J$ is a generalization of the polarity of each memristor. In general, a change of current orientation can in fact reabsorbed in this matrix. In the following, we set $\mathcal J=I$ for simplicity.  The memristor variables $w_i$ enter the dynamics by means of the diagonal matrix $W=\text{diag}(\vec W)=\text{diag}(w_1,\cdots,w_n)$, where with $\text{diag}(\vec \cdot)$ we mean the diagonal matrix with the input vector as diagonal elements. By definition, barred quantities ($\bar \Omega$, $\bar S$) depend on $P$. As stressed in \cite{{Caravelli2016rl}}, these equations are true in the \textit{bulk}, e.g. when all the memristors are not close to $w=\{0,1\}$, and has been derived assuming $W$ invertible. 
This is in general not true close to the lower boundary $w=0$, but in simulations we have observed a smooth behavior also in such case \cite{Caravelli2016}. This implies that together with the equations (\ref{eq:exacteqc}), one should impose the constraint $0\leq w_i \leq 1$ to have a faithful dynamics of a physical memristive system. 

Of course, eqn. (\ref{eq:exacteq}) equation describes a specific class of memristors (ideal memristors) and is by no means general. It can however be interpreted as first approximations for the real dynamics of memristors with a pinched hysteresis loop, in which the resistance is linear in the internal parameter, and the internal parameter varies linearly in the current.

As we have seen in Sec. \ref{sec:graphs}, $A$ and $B^t$ are dual, i.e. they satisfy $A^t B = B^t A=0$.
In general, for \textit{any} projector operator $\Omega=A^t(  A  A^t)^{-1} A$, if $A B^t=A^t B=0$, then it is possible to show
that 
\begin{equation}
I-A^t(  A  A^t)^{-1} A=I-\Omega=B(  B^t  B)^{-1} B^t,
\label{eq:dual}
\end{equation}
e.g. the fact that $\Omega$ can be written both in terms of $B$ or $A$.
Given the fact that $B$ is numerically much easier to calculate than $A$, this turns useful when performing simulations.

Equation (\ref{eq:exacteqc}) may seem quite obscure at first. Below we provide an example where calculations can be made without approximations. We show that known results can be re-derived using the equations above.\ \\\ \\ 
\textbf{A specific example: exact mean-field solution for memristors series.}

In order to see how the equations above can be applied, let us consider a simple enough case in which calculations can be performed analytically, and which are nonetheless not trivial: this is the case of a series of $N$ current-controlled memristors, for which in principle eqn. (\ref{eq:exacteqc}) would not be needed. In this case the use of the graph-theoretical machinery is an overkill which however provides insights in the meaning of the operator $\Omega$.

For a series of memristors, the adjacency matrix is a Toeplix matrix. Thus, the projector operator on the incidence matrix can be written as $\Omega_B=1-\frac{1}{N} \vec u \vec u^t$, where $\vec u=[1 \cdots 1]^t$ is a column vector of size $N$ with only $1$'s. Thus, $\Omega=1/N \vec u \vec u^t$. We can thus calculate the inverse $(1+\xi \Omega W)$ if $W$ has the same entries. In fact, we can use in this case the Sherman-Morrison identity \cite{smid}: one has that for any $k\in\mathbb{R}$, $(I+ k \vec u \vec u^t)^{-1}=I-\frac{k}{1+kN} \vec u \vec u^t$, thus if all memristors have the same initial value, one has $(1+\xi w \Omega)^{-1}=(1+\frac{\xi w}{N} \vec u \vec u^t)^{-1}=I-\frac{\xi w}{N(1+\xi w)} \vec u \vec u^t$. We can calculate the rhs of the dynamical equations exactly:
\begin{eqnarray}
\frac{d}{dt} w(t) \vec 1&=&-\frac{1}{\beta}(I-\frac{\xi w}{N(1+\xi w)} \vec u \vec u^t)\frac{1}{N}(\vec u \vec u^t) \vec s \nonumber \\
&=&-\frac{1}{\beta}(I-\frac{\xi w}{(1+\xi w)} ) \frac{1}{N} \vec u \vec u^t\ \vec s \nonumber \\
&=&-\frac{1}{\beta}\frac{1}{(1+\xi w)}  \frac{1}{N} \vec u \vec u^t\ \vec s.
\end{eqnarray}
We note that $\frac{1}{N}\vec u \vec u^t$ is a projector, which implies:
\begin{equation}
\frac{d}{dt} w(t) =-\frac{1}{ \beta}\frac{1}{(1+\xi w(t))}  \frac{1}{N}s(t).
\label{eq:finalex}
\end{equation}
Eqn. (\ref{eq:finalex}) is the same that would be obtained for a series of $N$ identical memristors if there was only a source voltage which is homogeneous across the circuit. Using the fact that we are approximating all the memristors with the same parameter, we have use the rule for the series of resistors, to obtain:
\begin{eqnarray}
I&=&\frac{V}{R}\equiv \frac{s(t)}{\sum_i R_i(w)} \equiv \frac{s(t)}{\sum_i \left( R_{on} w+(1-w) R_{off} \right)} \nonumber \\
&=&\frac{s(t)}{N \left( R_{on} w+(1-w) R_{off} \right)}=\frac{1}{N R_{on}} \frac{s(t)}{\left(1+\xi w\right)}
\end{eqnarray}
and using the fact that $\frac{d}{dt} w=\alpha w-\frac{R_{on}}{\beta }I $ we obtain the final equation (\ref{eq:finalex}). We note that if $\vec w$ is not uniform (i.e. when we do not use the mean field approximation), then it is not possible to neglect the correlation arising from the denseness of $\Omega$.

\subsection{Network constraints and effective memory}
Network constraints are fundamental in order to make precise the notion of \textit{effective memory} in memristive networks. In order to see this, let us look at the constraints and how these affect the internal memory capacity. The analysis which follows below applies to the case of a \textit{linear} relationship between the memristor' memory and either voltage or current. 
We consider first two specific models for the evolution of the internal memory in the ``bulk" (far from the boundaries). These are
\begin{equation}
\text{Current-Controlled memristors:}\ \ \ \frac{d}{dt} w=a \vec i,
\label{eq:currentcontr}
\end{equation}
and
\begin{equation}
\text{Voltage-Controlled memristors:}\ \ \  \frac{d}{dt} w=b \vec v,
\label{eq:voltagecontr}
\end{equation}
which are two different types of memristors considered in the literature \cite{qmemr,physpropr}, and $a$ and $b$ are simply constants.
In both eqns. (\ref{eq:currentcontr}) and (\ref{eq:voltagecontr}) one can uses the network constraints in order to obtain information on the exact number of independent memory states stored by the circuit.  In order to see this, we note first that the Kirchhoff current constraint can be written as:
\begin{equation}
B \vec i =0,
\end{equation}
and if we now combine the internal memory equation for the current-controlled memristors of eqn. (\ref{eq:currentcontr}), it is easy to see that:
\begin{equation}
B \vec i \propto B \frac{d}{dt} \vec w = 0.
\end{equation}
At this point we  can use the  tree and co-tree splitting to write the following linear relationship:
\begin{equation}
B \frac{d}{dt} \vec w = B_{\mathcal T} \frac{d}{dt} {\vec w}_{\mathcal T}+B_{\bar {\mathcal T}}  \frac{d}{dt}{\vec w}_{\bar {\mathcal T}}  =0.
\end{equation}
We thus obtain the final formula which connects the derivatives of the memory on the tree and the co-tree elements:
\begin{equation}
  \frac{d}{dt} {\vec w}_{\mathcal T}= - B_{\mathcal T} ^{-1}B_{\bar {\mathcal T}} \frac{d}{dt} {\vec w}_{\bar {\mathcal T}}.
\end{equation}
This equation can be then integrated in time to obtain, for current-controlled memristors, the following result:
\begin{equation}
  {\vec w}_{\mathcal T}(t)= - B_{\mathcal T} ^{-1}B_{\bar {\mathcal T}}  \left( {\vec w}_{\bar {\mathcal T}}(t) +{\vec w}_{\bar {\mathcal T}}^0\right),
\label{eq:memorycurrent}
\end{equation}
where the constant ${\vec w}_{\bar {\mathcal T}}^0$ arises from integrating the equation over time.
Before we provide an interpretation of the above result, we want first to show that such analysis applies also in the case of voltage controlled memristors. In this case, rather than the conservation of the currents at the nodes, we use the conservation of voltage across a cycle. This can be represented by the following equation dependent this time on the cycle matrix $A$:
\begin{equation}
A \vec v \propto A \frac{d}{dt} \vec w = 0.
\end{equation}
If we use the tree-chords splitting again, we have the same formalism as before, by replacing $B_{\cdot {\mathcal T}}$ with $A_{\cdot {\mathcal T}}$. We thus obtain:
\begin{equation}
{\vec w}_{\mathcal T}(t) =-A _{\mathcal T} ^{-1}A _{\bar {\mathcal T}}\left( {\vec w}_{\bar {\mathcal T}}(t) +{\vec w}_{\bar {\mathcal T}}^0\right).
\label{eq:memoryvoltage}
\end{equation}
Both equations (\ref{eq:memorycurrent}) and (\ref{eq:memoryvoltage}) are representations of the constraints of the network.  In both cases, we can write the equation for the internal memory as:
\begin{equation}
{\vec w}_{\mathcal T}(t)= Q_{\mathcal T} \left( {\vec w}_{\bar {\mathcal T}}(t) +{\vec w}_{\bar {\mathcal T}}^0\right)
\end{equation}
where $Q_{\mathcal T}$ is a linear operator which depends on the chosen spanning tree, and thus for the whole memory as
\begin{eqnarray}
\vec w(t) &=&(Q_{\mathcal T}  {\vec w}_{\bar {\mathcal T}}(t),  {\vec w}_{\bar {\mathcal T}} (t))+(Q_{\mathcal T} {\vec w}_{\bar {\mathcal T}}^0,{\vec 0}_{\bar {\mathcal T}})\nonumber\\
&=&(Q_{\mathcal T}  , I  ){\vec w}_{\bar {\mathcal T}} (t)+(Q_{\mathcal T} {\vec w}_{\bar {\mathcal T}}^0,{\vec 0}_{\bar {\mathcal T}})
\label{eq:affinerel}
\end{eqnarray}
This is general, and it is valid both for current-controlled and voltage-controlled memristors, as long as these are linear at the first order approximation. It is easy to see that eqn. (\ref{eq:affinerel}) establishes an affine relationship between the internal memory and a subspace of chord memory.  We can thus introduce the concept of \textit{effective} memory of a memristive circuit $\mathcal G$, given by:
\begin{equation}
\Gamma(\mathcal G)=\frac{|E(\mathcal G)|-|\mathcal T(\mathcal G)|}{|E(\mathcal G)|}
\label{eq:effcap}
\end{equation}
where $|E(\mathcal G)|$ is the number of memristive elements and $|\mathcal T(\mathcal G)|$ represents the cardinality of a maximal spanning tree in the circuit $G$. Since $M$ can grow as the number of nodes of the circuit square, meanwhile $|\mathcal T(\mathcal G)|$ grows linearly in the number of nodes, this implies that denser circuits can effectively overcome the limitation of a smaller internal capacity. We note that the effective capacity of eqn. (\ref{eq:effcap}) is well defined: this number is independent from the choice of the spanning tree, and thus is a relevant physical quantity, meanwhile eqn. (\ref{eq:affinerel}) implicitly depends on the choice of the spanning tree. Specifically, the number of ways in which eqn. (\ref{eq:affinerel}) can be written depends on the number of spanning trees of the circuit. 

As simple as such argument may look, it shows that the effective memory in a memristive circuit lives on a submanifold of the internal memory. Once a spanning tree has been chosen, and the dynamical equations derived, the projection operator ensures that such sub-manifold is protected and well defined in the dynamics.

\subsection{Strongly and weakly non-linear regimes: two different limits for the dynamics}
In this section we study the behavior of the dynamics in the weak and strong non-linear regimes.
There are at least two regimes that we would like here to describe: $\xi\approx0$, which we call \textit{weakly non-linear regime}, and $\xi\rightarrow \infty$, we call \textit{strongly non-linear regime}. We focus here on the case of current-controlled memristors, but a similar analysis applies also to voltage-controlled memristors. The differential equation for the internal dynamics, for $\alpha=0$, is given by:
\begin{equation}
\frac{d}{dt} \vec W=-\frac{1}{\beta} \left(I+\xi \bar \Omega W\right)^{-1} \bar \Omega \bar S=-\frac{1}{\beta} T(\xi) \vec S,
\label{eq:tomultiply}
\end{equation}
where we introduced the definition of the operator $T(\xi)$.
The two regimes of $\xi$ can be understood from the analysis of the behavior of the operator 
\begin{equation}
T(\xi)\equiv\left (I+(\frac{R_{on}}{R_{off}}-1)\ \bar \Omega W\right)^{-1}\bar \Omega=\left(I+\xi\ \bar \Omega W\right)^{-1}\bar \Omega,
\end{equation}
which we will now try to make precise from an operatorial point of view in both limits.

In the weakly nonlinear regime, i.e. $R_{off}\approx R_{on}$, the following Taylor expansion of the operator applies:
\begin{equation}
\lim_{\xi\rightarrow0}T(\xi)\approx\left(I-\xi\ \bar \Omega W\right) \bar \Omega+O(\xi^2).
\end{equation}
This regime will be studied in detail in Sec. \ref{sec:learning}, showing that we can identify the ``learning" abilities of the memristive circuit.  
In the strong non-linear regime instead, it does make sense to write
\begin{equation}
\lim_{\xi\rightarrow\infty}T(\xi)=\lim_{\xi\rightarrow\infty} \frac{1}{\xi} \left(\frac{1}{\xi}I+\bar \Omega W\right)^{-1}\bar \Omega.
\end{equation}

We note that $\left(\frac{1}{\xi}I+\bar \Omega W\right)^{-1}$ for large $\xi$, can be seen as the Tychonov regularization of the inverse of the operator $\bar \Omega W$. Eqn. (\ref{eq:exacteqc}) was derived with the assumption that $W$ is an invertible (diagonal) matrix, i.e. that no memristor reached the $R_{on}$ state. 
We can thus write 
\begin{equation}
\lim_{\xi\rightarrow\infty} \left(\frac{1}{\xi}I+\bar \Omega W\right)^{-1}\left(\bar \Omega W\right) (\xi W)^{-1}, 
\label{eq:tobem}
\end{equation}
and study for the time being how does $\lim_{\xi\rightarrow\infty} \left(\frac{1}{\xi}I+\bar \Omega W\right)^{-1}\left(\bar \Omega W\right)$ behave. The Tychonov regularization converges to the Moore-Penrose pseudo-inverse of $\bar \Omega W$, implying that $\lim_{\xi\rightarrow\infty} \left(\bar \Omega W\right) \left(\frac{1}{\xi}I+\bar \Omega W\right)^{-1}\left(\bar \Omega W\right)=\bar \Omega W$. If we multiply eqn. (\ref{eq:tomultiply}) on the left by $\bar \Omega W$, we can write:
\begin{eqnarray}
\lim_{\xi\rightarrow \infty}\bar \Omega W \frac{d \vec W}{d t}&=&-\lim_{\xi\rightarrow \infty} \frac{1}{\xi \beta}\left(\bar \Omega W\right) \left(\frac{1}{\xi}I+\bar \Omega W\right)^{-1}\left(\bar \Omega W\right) W^{-1} \bar S\nonumber \\
&=&-\frac{1}{\xi \beta} \bar \Omega W W^{-1} \bar S =-\frac{1}{\xi \beta} \bar \Omega \bar S.
\end{eqnarray} 
From the equation above we derive a conservation law by integration over time, in the limit $\xi\gg1$:
\begin{equation}
\bar \Omega \left[\ \frac{\xi \beta}{2}\left(\vec {W}^2(t)-\vec {W}^2(t_0)\right)+\int_{t_0}^t \bar S(\tilde t) d\tilde t\ \right]=0,
\end{equation}
where $\vec {W}^2$ means the vector with all the elements squared. In general, it is easy to see that this equation is true up to an arbitrary vector $\vec k(t)$, obtaining
\begin{equation}
\frac{\xi \beta}{2}\left(\vec {W}^2(t)-\vec {W}^2(t_0)\right)+\int_{t_0}^t \bar S(\tilde t) d\tilde t\ +(I-\bar \Omega) \vec k(t)=0,
\label{eq:conslaw}
\end{equation}
which is the final conservation law in this limit, similar to what observed in \cite{galeconv}.  Eqn. (\ref{eq:conslaw}) is true only in the approximation in which the dynamics lies in the bulk, i.e. $0< W_i(t)<1$. Also, we note that in order to derive this conservation law, we needed to introduce $W^{-1}$, which is not invertible if some memristors are in the state $W_i=0$, but the final result is independent from the inverse.

\section{General properties of the dynamics}
\subsection{Symmetries and dualities}
Equation (\ref{eq:currentcontr}) satisfies several symmetries which we would like here to describe in detail. Let us first start by saying that the dynamical equations obtained depend on the choice of a spanning tree to begin with: the operator $\bar \Omega$ should in fact be more correctly written as $\bar \Omega_{\mathcal T}$ to be precise. The results we obtain do not depend on the choice of the tree $\mathcal T$, but the equations do. This is an example of a gauge degree of freedom. In addition, the equations depend on the choice of a direction of the currents on the circuit. However, these can be reabsorbed in the matrix $\mathcal J$ introduced before, as these are simply the signs associated to the current.
A general $\mathcal Z_2$ symmetry is however easier to see for a global change of current signs: under a change of current direction, $A \rightarrow -A$. We note that $\bar \Omega$ is independent from this transformation, as it depends on an even number of matrices $A$. Another symmetry of the dynamics is given by a change of active components to passive components and viceversa. Formally, this implies $P\rightarrow -P$: again, since $P$ appears twice in $\bar \Omega$, the dynamics is unchanged.

Another symmetry to be expected is the transformation $\vec S \rightarrow - \vec S$ and $t\rightarrow -t$, which reverses voltages and time, this however for the specific case of $\alpha=0$.

Let us now consider a linear transformation of the $w_i(t)$ involved, i.e. ${\vec W}^\prime(t)=O^{-1} \vec W(t)$, where O is an invertible matrix. In this case, $W(t)=\text{diag}\left(\vec W(t)\right)\rightarrow W^\prime(t)=O^{-1} \text{diag}\left( W(t) \right) O$. In order to see this, let us 
look at how the equation transforms under a change of basis for $W$. We first note that $\frac{d {\vec W}^\prime}{dt}=O^{-1}\frac{d \vec W}{dt}$. Thus:
\begin{eqnarray}
\frac{d {\vec W}^\prime}{dt}&=&O^{-1}\frac{d \vec W}{dt}\nonumber \\
&=&\alpha  O^{-1} \vec W-\frac{1}{\beta}O^{-1}\left(I+\xi \bar \Omega W\right)^{-1} O O^{-1} \bar \Omega O O^{-1} \bar S \nonumber \\
&=&\alpha  {\vec W}^\prime-\frac{1}{\beta}O^{-1}\left( I +\xi  \bar \Omega O O^{-1} W \right)^{-1} O O^{-1}  \bar \Omega O O^{-1}  \bar S \nonumber  \\
&=&\alpha  {\vec W}^\prime -\frac{1}{\beta}\left( I +\xi   {\bar \Omega}^\prime ( O^{-1} W O) \right)^{-1}   {\bar \Omega}^\prime  {\bar S}^\prime,
\label{eq:transf}
\end{eqnarray}
where we defined ${\bar \Omega}^\prime=O^{-1} \bar \Omega O$ and ${\bar S}^\prime=O^{-1}  \bar S$. This shows for instance that if we choose a basis in which $W=\text{diag}(\vec W)$, then $\bar \Omega$ will not be diagonal. If on the other hand we choose a basis in which $\bar \Omega$ is diagonal, as a result $W$ will likely not be diagonal, unless $\bar \Omega$ and $W$ commute. One thing that needs to be stressed is that $O^{-1} W O \neq \text{diag}({\vec W}^\prime)=W^\prime(t)$. If however $O$ is a matrix which represents a specific permutation of the memristive element labels, then $W^\prime(t)$ will still be diagonal with the elements on the diagonal permuted accordingly.

One feature which becomes clear in eqn. (\ref{eq:currentcontr}), is the fact that not all components of the source vector affect the evolution of the internal memory. In fact, we could add an arbitrary vector $\Delta S=(I-\bar \Omega) \vec k$ to $\bar S$: since $\bar S$ enters the equation as $\bar \Omega \bar S$, one automatically has that 
\begin{equation}
\bar \Omega (\bar S+\Delta S)=\bar \Omega \bar S+ \bar \Omega (I-\bar \Omega) \vec k=\bar \Omega \bar S.
\end{equation}
This is a result which is reminiscent of the network constraints, or alternatively interpreted as a gauge freedom. Thus, we can easily decompose $\bar S$ using the identity $I=\bar\Omega + (I-\bar \Omega)$, as 
\begin{equation}
\bar S=I \bar S=\bar \Omega \bar S+(I-\bar \Omega) \bar S,
\end{equation}
where the second term on the right hand side is in the orthogonal subspace respect to $\bar \Omega$, as $\bar \Omega (I-\bar \Omega)=0$. Since in eqn. (\ref{eq:exacteq}) we have that the vector $\bar S$ has the matrix $\bar \Omega$ applied on the left, the components of $\bar S$ orthogonal to $\bar \Omega$ do not contribute to the dynamics. 
This result is important in light of the fact that the applicability of memristive circuits to, for instance, machine learning, depends on the ability to control the dynamics by means of external voltages.

\subsection{Passive/Active components and oscillations}
In this section we wish to show that oscillations around an asymptotic fixed point can be present only when  $\bar \Omega$ is not symmetric. 
By construction, this happens when $P\neq \pm I$, being $\bar \Omega=A^t( \bar A A^t)^{-1} \bar A$. Physically, this represents the case in which only a mixture of active and passive components are present in the circuit (which are in our formalism represented by positive and negative resistances).

First we work out a simple exercise which will turn out to be useful later. Specifically, this will be in the case $P=\pm I$, for which $\bar \Omega\rightarrow \Omega$. One key element of the proof which follows below is the analysis of matrix similarity, for which a matrix has similar eigenvalues to another matrix. For instance, although $ \Omega W$ is not a symmetric matrix, it has always real eigenvalues. In order to see this, we note that the eigenvalues of any matrix product $\bar \Omega W$ has the same eigenvalues of the matrix $W^{\frac{1}{2}} \bar \Omega W^{\frac{1}{2}}$. In this case, since $W$ is diagonal and positive, the square root of the matrix is simply the square root of the diagonal elements. This is due to the fact that
any matrix $Q M Q^{-1}$, for any invertible matrix $Q$, has the same eigenvalues as those of $M$. In the following, we use the symbol $\sim$ for similarity, i.e. matrices with similar eigenvalues ($M\sim Q M Q^{-1}$). If $\bar \Omega$ is symmetric and real, then $W^{\frac{1}{2}} \bar \Omega W^{\frac{1}{2}}$ has real eigenvalues as it is a symmetric matrix. This implies that also $(I+\xi \bar \Omega W)^{-1}$ has real eigenvalues. In fact, we have that 
$$(I+\xi \Omega W)^{-1}\sim W^{\frac{1}{2}} (I+\xi  \Omega W)^{-1} W^{-\frac{1}{2}}=(I+\xi W^{\frac{1}{2}}  \Omega W^{\frac{1}{2}})^{-1}.$$ Since the inverse of a symmetric matrix is symmetric, again its eigenvalues must be real.  Also, since $\bar \Omega$ is invariant under the transformation $P\rightarrow -P$, such analysis applies also for the inverse system, in which the number of passive and active component has been exchanged.

On the other hand, this is \textit{not} true if $\bar \Omega$ is not symmetric, and thus $\bar \Omega W$ can have pairs of complex eigenvalues. Let us now assume that the spectrum of $\bar \Omega W$ is $\left(\lambda_1, \cdots, \lambda_n, \lambda_{n+1},\lambda_{n+1}^*, \cdots,  \lambda_{n+k},\lambda_{n+k}^*\right)$.  Then, the spectrum of $R=\left(I+\xi \bar \Omega W\right)^{-1}$ will be of the form
\begin{equation}
\sigma(R)=\left((1+\xi \lambda_1)^{-1}, \cdots, (1+\xi \lambda_n)^{-1}, \frac{1+\xi \lambda_{n+1}^*}{(1+\xi \lambda_{n+1}^*)(1+\xi\lambda_{n+1})},\frac{1+\xi \lambda_{n+1}}{(1+\xi\lambda_{n+1}^*)(1+\xi \lambda_{n+1})}, \cdots\right) 
\end{equation}
and thus still possibly contains pairs of complex eigenvalues.   This observation is key to show that in the case in which $\bar \Omega$ is symmetric (i.e. no mixture of active/passive components) there cannot be oscillations around fixed points.

In order to see this, let us now consider the linearized dynamics close to a fixed point $W^*$,
\begin{equation}
\frac{d}{dt} \vec W\approx L|_{{\vec W}^*} \vec W,
\end{equation}
where $L|_{{\vec W}^*}$ is the Lyapunov matrix, given by
\begin{eqnarray}
L_{ji}&=&\partial_{w_i} f_j(\vec w)|_{{\vec W}^*} \nonumber \\
&=&\partial_{w_i} \sum_k \left(  (I+\xi \bar \Omega W)^{-1} _{j k} (\bar \Omega \bar S)_k\right)|_{{\vec W}^*} \nonumber \\
&=&\partial_{w_i} \sum_k \left(  \partial_{w_i} (I+\xi \bar \Omega W)^{-1} _{j k}\right) (\bar \Omega \bar S)_k|_{{\vec W}^*}.
\label{eq:lyapunov}
\end{eqnarray}
In the calculations which follow we will use the formula $\partial_{w_i} W_{jk}=\partial_{w_i} \left(w_j \delta_{jk}\right)=\delta_{ij}\delta_{jk}$, and the formula $\partial_s A^{-1}=-A^{-1} (\partial_s A) A^{-1}$ for any scalar quantity $s$. We thus have:
\begin{eqnarray}
L_{ji}(W^*)&=&-\xi \sum_{k_1,k_2,k_3,k_4}   (I+\xi \bar \Omega W)^{-1} _{j k_1}\bar \Omega_{k_1 k_2} \left(\partial_{w_i} W_{k_2 k_3}  \right)(I+\xi \bar \Omega W)^{-1} _{k_3 k_4}(\bar \Omega \bar S)_{k_4}|_{{\vec W}^*} \nonumber \\
&=&-\xi \sum_{k_1,k_2,k_3,k_4}   (I+\xi \bar \Omega W)^{-1} _{j k_1}\bar \Omega_{k_1 k_2}  \delta_{i k_2} \delta_{k_2 k_3}(I+\xi \bar \Omega W)^{-1} _{k_3 k_4}(\bar \Omega \bar S)_{k_4}|_{{\vec W}^*} \nonumber \\
&=&-\xi    \left( (I+\xi \bar \Omega W^*)^{-1}  \bar \Omega \right)_{j i} \sum_{k} \left((I+\xi \bar \Omega W^*)^{-1} \bar \Omega \right)_{i k} {\bar S}_{k},
\label{eq:lyapunov2}
\end{eqnarray}
which is a rather complicated expression. In the first line of eqn. (\ref{eq:lyapunov2}) we have used the derivative for the inverse, in the second the identity for the derivative of the diagonal matrix $W$, and in the third line we have simply summed over the indices and renamed the remaining indices. We wish to understand now what are the conditions for which the matrix $L$ above has only real eigenvalues (in which case no oscillations occur). This task can be achieved by showing that the matrix of eqn. (\ref{eq:lyapunov2}) is similar to a real symmetric operator. First we note that also the matrix $L$ is of the form:
\begin{equation}
L=M D
\end{equation}
where $M$ is a full matrix and $D$ is non-zero only on the diagonal. We do not consider any restriction on the elements of $D$: these can either be positive or negative for the time being. We assume however that the fixed points $W^*$ are such that $w^*_i\neq0$.
The diagonal elements of $D$ are the vector elements $(I+\xi \bar \Omega W^*)^{-1} \bar \Omega  {\bar S}$ and are real, meanwhile $M= (I+\xi \bar \Omega W^*)^{-1}  \bar \Omega$. 
First, we write:
\begin{eqnarray}
L&=&{W^*}^{-\frac{1}{2}}{W^*}^{\frac{1}{2}}(I+\xi \bar \Omega W^*)^{-1} {W^*}^{-\frac{1}{2}}{W^*}^{\frac{1}{2}}\bar \Omega W^{\frac{1}{2}} W^{-\frac{1}{2}} D \nonumber \\
&=&{W^*}^{-\frac{1}{2}}(I+\xi {W^*}^{\frac{1}{2}}\bar \Omega {W^*}^{\frac{1}{2}} )^{-1}{W^*}^{\frac{1}{2}}\bar \Omega {W^*}^{\frac{1}{2}} {W^*}^{-\frac{1}{2}} D \nonumber \\
&=& {W^*}^{-\frac{1}{2}}  (I+\xi X )^{-1} X  {W^*}^{-\frac{1}{2}} D=M D\sim \sqrt{D} M \sqrt{D}
\label{eq:lyapunov3}
\end{eqnarray}

 In the first line we have used the identity $I={W^*}^{-\frac{1}{2}}{W^*}^{\frac{1}{2}}$. The square root matrix exists for positively defined matrices, such as the diagonal matrix $W^{*}$ which, by construction, is invertible. In the second line we have used the identity  ${W^*}^{\frac{1}{2}}(I+\xi \bar \Omega W^*)^{-1} {W^*}^{-\frac{1}{2}}=(I+\xi {W^*}^{\frac{1}{2}}\bar \Omega {W^*}^{\frac{1}{2}} )^{-1}$. In the third line we have implicitly introduced the definition $X={W^*}^{\frac{1}{2}}\bar \Omega {W^*}^{\frac{1}{2}}$. We have also implicitly defined the matrix $M={W^*}^{-\frac{1}{2}}  (I+\xi X )^{-1} X  {W^*}^{-\frac{1}{2}}$.
\ \\
We now observe that $(I+\xi X )^{-1} X$ is symmetric if and only if also $X$ is symmetric. In order to see this, we use the identity $\left((I+\xi X )^{-1} X\right)^t= X^t (I+\xi X^t )^{-1}$;  we note that the identity $\left((I+\xi X )^{-1} X\right)^t= X (I+\xi X )^{-1}$ holds due to the fact that $X$ commutes with itself. Thus, the matrix $M$ is symmetric if and only if $X$ is symmetric. In general, we have that $MD$ has the same eigenvalues of $\sqrt{D}M\sqrt{D}$. This implies that, if $M$ is symmetric, the eigenvalues of $MD$ are real if and only if $\sqrt{D}$ is real. We can thus state that, given the fixed points of the dynamics $W^*(\bar S)$, there are no oscillations if the following two conditions are satisfied:
\begin{enumerate}
\item $X$ (and thus $\bar \Omega$) is symmetric and 
\item $D_{ii}=\sum_{k} \left(\left(I+\xi  \Omega W^*(\bar S)\right)^{-1}  \Omega \right)_{i k} {\bar S}_{k} >0\ \forall i$.
\end{enumerate}
All these facts put together show that $L$ is similar to a real symmetric operator (and thus with real eigenvalues) if and only if the conditions above are satisfied. Let us note that in the proof we have not chosen a specific fixed point $W^*$, although the condition (2) depends implicitly only the source vector $\vec S$ and the topology of the circuit. Since $\bar \Omega$ is symmetric only if $P=\pm I$, this proves a weak result, e.g. the fact that is a constraint on the presence of oscillations around fixed points.
In order to make this result stronger, we need to specify whether $D_{ii}$ are effectively non-negative at the fixed point. In order to do that, we need to use the equations of motion and the fixed points structure. This is given by:
\begin{equation}
0=\frac{d}{dt} \vec W(t)=\alpha \vec W(t)-\frac{1}{\beta}\left(I+\xi \Omega W \right)^{-1} \Omega \vec S
\end{equation}
from which we obtain the fact that at the fixed point:
\begin{equation}
\alpha \vec W^*=\frac{1}{\beta}\left(I+\xi \Omega W^* \right)^{-1} \Omega \vec S.
\end{equation}
We now notice that on the right hand side we have a vector which is the definition of the diagonal elements of $D_{ii}$. We can thus now confidently say that $D_{ii}=\alpha \beta W^{*}$. Since $\alpha>0$ and $\beta>0$, and $1>W^*>0$ by construction, this completes the proof that we had anticipated.
 Analogously to the \textit{Barkhausen criterion} for circuits with feedback loops \cite{barkhausen}, this condition is necessary but not sufficient, i.e. there could be memristive circuits with passive/active elements mixtures which do not have oscillations. 




\section{Dynamics as a constrained gradient descent optimization} \label{sec:learning}

In the previous sections we have studied the properties of the dynamics of purely memristive circuits.
In this section we aim to look at the dynamics from another angle, which is spirit close to self-organizing maps as described long ago by Kohonen \cite{Kohonen}. A precise statement which connects memristors to an optimization problem will be made in what follows. We first  consider one specific case as a warm up: the mean field problem in which we use the ansatz $\vec W = w(t) \vec 1$, and for the case of only passive (or active) components. In this case, the vector
\begin{equation}
-\frac{1}{\beta}\left(I + \xi w(t)  \Omega \right)^{-1} \Omega \bar S,
\label{eq:solreg}
\end{equation}
which appears on the right hand side of the differential equation in (\ref{eq:exacteqc}), can be interpreted as a tentative linear regression. Let us in fact assume that we aim to solve the equation:
\begin{equation}
 \Omega \vec r(t) = \bar S(t)
\end{equation}
given a specific regularizer \cite{Golub} for the variable $\vec r(t)$.
Such  (ill-posed) equation can be solved by means of a Tychonov regularization:
\begin{equation}
\text{min}_{\vec r(t)}\ \ ||\Omega \vec r(t)-\bar S(t)||^2+ || \frac{1}{\sqrt{\xi w(t)}} \vec r(t)||^2
\end{equation}
where the norm $||\cdot||^2$ is the standard $L_2$ vector norm, which gives 
\begin{equation}
\vec r^*=\left(\frac{1}{\xi w(t) }I +  \Omega^t \Omega \right)^{-1} \Omega \bar S(t).
\end{equation}
Since $\Omega$ is symmetric and a projector, one has
\begin{equation}
\vec r^*=\left(\frac{1}{\xi w(t) }I +  \Omega \right)^{-1} \Omega \bar S(t),
\end{equation}
which is proportional to the vector of eqn. (\ref{eq:solreg}). The factor $\vec r^*$ is interpreted as the time derivative of the internal state $w(t)$ in the homogeneous case. The internal state thus moves in the direction which minimizes the least square problem above.
This result hints toward the fact that memristive systems are performing a specific type of optimization.
However, in the general case things are slightly more complicated and at the moment we do not have a complete answer of what type of optimization these systems are performing.  Notwithstanding these difficulties, there is something we can say in weak non-linear regime, $\xi\ll 1$. We want to interpret eqn. (\ref{eq:exacteqc}) as a specific dynamics which is of interest to machine learning, and in general to optimization problems. For simplicity, we consider the case $P=I$ and $\alpha=0$. Specifically, let us consider the following time-discretized dynamics:
\begin{equation}
 \vec W(t+1)= \vec W(t)-\frac{dt}{\beta}\left(I+\xi  \Omega W(t)\right)^{-1}  \Omega  \vec S(t)  \\
\end{equation}
which, in the approximation $\xi\ll 1$ can be written as
\begin{eqnarray}
 \vec W(t+1)&\approx&\vec W(t)-\frac{dt}{\beta}\left(I-\xi  \Omega W(t)\right) \Omega  \vec S  (t)\nonumber\\
 &=&\vec W(t)-\frac{dt}{\beta}\left(I-\xi  \Omega \right) \text{diag}\left( \Omega  \vec S(t)\right) \vec W(t)  \nonumber.
\end{eqnarray}
We now use the fact that if $ \Omega$ is projector, then one can use the identity
\begin{equation}
 \Omega\ \text{diag}\left( \Omega \vec S(t)\right)=\text{diag}\left(  \Omega \vec S(t)\right),
\end{equation}
from which we can derive:
\begin{eqnarray}
 \vec W(t+1)&\approx&\vec W(t)+\frac{dt(\xi-1)}{\beta}  \Omega\   \text{diag}\left(\Omega \vec S(t)\right) \vec W(t).
 \label{eq:discretedyn}
\end{eqnarray}
The equation above can be rewritten as:
\begin{eqnarray}
 \vec W(t+1)&\approx& \vec W(t)+\mu   \Omega\ \vec \nabla_{\vec W} f(\vec W),
\end{eqnarray}
where we defined $\frac{dt(\xi-1)}{\beta} \equiv \mu$ and $\vec \nabla_{\vec W} f(\vec W)=\text{diag}\left( \Omega  \vec S(t)\right) \vec W(t)$.
Now that we have written the dynamical equation in this fashion, it is easy to realize that the dynamics is effectively a gradient descent procedure for a constrained optimization problem. We claim that such dynamics performs a constrained optimization of the type:
\begin{eqnarray}
\text{minimize}&\ &f(\vec W)\ \text{s.t.} \\
&\ &B\vec W=0 
\end{eqnarray}
where $f(W)=\sum_{i,j}\frac{1}{2}  \Omega_{ij} S_j W_i^2$ and $B$ is the directed incidence matrix. 
In order to see this, let us consider Rosen's gradient projection method to solve this optimization problem \cite{Rosen}. The basic assumption of the gradient projection method is that $\vec W$ lies in the tangent subspace of the constraints. In order to provide an exact mapping, we consider first a general update rule given by:
\begin{equation}
\vec W_{t+1}=\vec W_t + \alpha \vec z
\end{equation}
where both $\vec W_{t+1}$ and $\vec W_t $ satisfy are assumed to satisfy the linear constraint, and which depends on an arbitrary vector $\vec z$. We restrict our attention to the case in which the vector $\vec z$ is in the steepest descent direction, and also satisfies $B \vec z=0$.  This condition ensures that if $\vec W_0$ satisfies the linear constraint, then $\vec W_t$ will also $\forall t>0$. To be clear, the goal is to show the equivalence between the discrete dynamics of eqn. (\ref{eq:discretedyn}) and the following optimization procedure:
\begin{eqnarray}
\text{minimize}&\ &\vec z \cdot \vec \nabla_{\vec W} f(\vec W)\  \\
&\ &\text{s.t.}\ \ B\vec z=0\ \text{and} \\
&\ &\vec z \cdot \vec z=1.
\end{eqnarray}
where $\vec \nabla_{\vec W} f(\vec W)=\left(\partial_{w_1} f(\vec W),\cdots,\partial_{w_M} f(\vec W)\right)$. 
We now follow the procedure introduced by Rosen in \cite{Rosen}. We introduce the Lagrange multipliers $\vec \lambda$ and $\mu$, and the Lagrangian:
\begin{equation}
\mathcal L (\vec z,\vec \lambda,\mu)=\vec z \cdot \vec \nabla_{\vec W} f(\vec W)-\vec s \cdot B \vec \lambda-\mu (\vec z \cdot \vec z-1).
\end{equation}
The Euler-Lagrange equations for $\vec s$ are given by:
\begin{equation}
\partial_{\vec z} \mathcal L =\vec \nabla_{\vec W} f(\vec W)- B \vec \lambda-2 \mu \vec z =0.
\label{eq:condition}
\end{equation}
If we multiply this equation by $B^t$ on the left hand side, we obtain the equation:
\begin{equation}
B^t \vec \nabla_{\vec W} f(\vec W)- B^t B \vec \lambda =0,
\end{equation}
from which we can invert for the Lagrange multiplier $\vec \lambda$:
\begin{equation}
\vec \lambda=(B^t B )^{-1}B^t \vec \nabla_{\vec W} f(\vec W),
\end{equation}
and thus using eqn. (\ref{eq:condition}) we finally obtain:
\begin{equation}
\vec z=\frac{1}{2\mu}\left(I- B(B^t B )^{-1}B^t\right) \vec \nabla_{\vec W} f(\vec W).
\end{equation}
Such vector can be now re-inserted into the dynamical equation, which is now interpreted as a constrained gradient descent:
\begin{equation}
\vec W_{t+1}=\vec W_{t}+\frac{1}{2\mu}(I- B(B^t B )^{-1}B^t) \vec \nabla_W f(\vec W)\equiv \vec W_{t}+\frac{1}{2\mu}\Omega \vec \nabla_{\vec W} f(\vec W).
\label{eq:gradientdescent}
\end{equation}
It is easy at this point to identify, a posteriori, every element in this equation. The projector operator is given by $ \Omega$. In the case in which only active or only passive element are present, we can use the duality between the loop matrix $A$ and the incidence matrix $B$, to write $\Omega=I- B(B^t B )^{-1}B^t$. Thus, the constraint $B \vec W=0$ can be interpreted exactly as the conservation of memory in the circuit, and $B$ promptly identified as the incidence matrix. The constant $\frac{1}{2\mu}\equiv\frac{dt(1-\xi)}{\beta}$ is also obtained, and all it is left to us to do is to identify $\left(\vec \nabla_W f(\vec W)\right)_i\equiv\sum_{j} \Omega_{ij}  S_j W_i$, from which after a simple integration we obtain the functional $f(\vec W)=\sum_{ij} \frac{\Omega_{ij}}{2}  S_j W_i^2$. 
This interpretation is key to identify memristive networks as ``learning": gradient descent is in fact one of main training algorithms in machine learning and optimization, and in particular in neural networks. Such connection establishes memristive circuits as the perfect neuromorphic devices. Of course, this is not the first time this was suggested \cite{memrnet}, but here we have provided further evidence of the above.  For instance, in \cite{carbajal} it was shown that in the case of a memristor series one can use the equations for learning. Using the fact that in the case of a series $\Omega=\frac{1}{N} \vec u \vec u^t$, and that $\sum_j u_j=N$, we can show that the functional being minimized was 
$f(\vec W)=\frac{1}{2N}\sum_{ij} S_i W_j^2$.

This is interesting also for other reasons. First of all, it makes precise the notion of information overhead for the specific case of purely memristive systems. In a recent paper, Di Ventra and Traversa \cite{traversa14b} put forward the suggestion that the graph topology is part of the optimization process. In the above we have just observed that the function being optimized is
$f(\vec W)=\sum_{ij} \frac{\Omega_{ij}}{2} S_j W_i^2$ 
in which both the external voltage sources \textit{and} the network topology (through $\Omega$) appears.
For the technological application of such statement, this poses the problem of engineering $\Omega$ and choose $\vec S$ in order to minimize the function of which one desires to find a minimum. 

\section{Conclusions}

In the present paper we have made several steps towards understanding the collective behavior of circuits with memory. We used a recently derived equation for the internal memory of an ideal and purely memristive system. Memristors, and in particular memristive circuits, are interesting devices with a very rich dynamical behavior. Even for the simpler memristor type (linear), non-linear phenomena emerge at the dynamical level. In fact, such an equation
 establishes that the non-linearity is controlled by a single parameter, which is the ratio between the resistance in the insulating phase and the resistance in the conducting phase of the memristor. Here we focused on the technical aspects and properties of the derived equation and tried to create a link between the dynamics of the internal memory to a more standard machine learning approach. Specifically, we have described in detail the symmetries of eqn. (\ref{eq:exacteq}), and analyzed the difference between purely passive (or active) systems and their mixtures. We have proven a condition on the existence of complex eigenvalues of the Lyapunov matrix. This was done by showing that the matrix which describes the dynamics close to a fixed point is similar to a real symmetric operator. We have shown that only for mixture of passive and active components one can have complex Lyapunov exponents.

We have also discussed what type of optimization purely memristive systems are performing using the internal memory equation. As it turns out, in the limit of weak non-linearity, the dynamical equation can be casted in the form of a (linear) constrained gradient descent equation \cite{Rosen}. The functional being minimized was found to be a combination of sources and quadratic in the memristor internal parameters, weighted by the projector operator on the space of cycles.  More complex optimizations require the introduction of other circuital elements \cite{traversa14b} which we did not consider in the present paper. 

\ \\\textbf{Acknowledgements.} We would like to thank Fabio Lorenzo Traversa and Massimiliano Di Ventra for various discussions concerning memristors. I would also like to thank the anonymous editors of a previous paper for saying that such equations were not useful. I would like to thank the anonymous referees and J.P. Carbajal for several comments which improved both the readability of this paper and spotting few incorrect equations.


\end{document}